\documentclass{rspublic}
\usepackage{euscript,amssymb,amsmath,harvard}
\usepackage[dvips]{graphicx}

\newcommand{\be}[1]{\begin{equation}\label{#1}}
\newcommand{\ee}{\end{equation}}
\newcommand{\ba}[1]{\begin{eqnarray}\label{#1}}
\newcommand{\ea}{\end{eqnarray}}
\newcommand{\rf}[1]{(\ref{#1})}
\newcommand{\nn}{\nonumber}

\title{Campbell diagrams of weakly anisotropic flexible rotors}

\author{{O.N. Kirillov}\footnote{Dynamics and Vibrations Group, Department of Mechanical Engineering,
Technical University of Darmstadt,
Hochschulstr. 1, 64289 Darmstadt, Germany (kirillov@dyn.tu-darmstadt.de).}}

\begin{document}
\maketitle

\begin{abstract}{Campbell diagram; flexible rotor; dissipation-induced instabilities;
subcritical flutter; symplectic (Krein) signature; non-Hermitian degeneracies}

We consider an axi-symmetric flexible rotor perturbed by dissipative, conservative, and non-conservative positional forces originated at the contact with the anisotropic stator. The Campbell diagram of the unperturbed system is a mesh-like structure in the frequency-speed plane with double eigenfrequencies at the nodes. The diagram is convenient for the analysis of the traveling waves in the rotating elastic continuum. Computing sensitivities of the doublets we find that at every particular node the unfolding of the mesh into the branches of complex eigenvalues in the first approximation is generically determined by only four $2\times2$ sub-blocks of the perturbing matrix. Selection of the unstable modes that cause self-excited vibrations in the subcritical speed range, is governed by the exceptional points at the corners of the singular eigenvalue surfaces---`double coffee filter' and `viaduct'---which are sharply associated with the crossings of the unperturbed Campbell diagram with the definite symplectic (Krein) signature. The singularities connect the problems of wave propagation in the rotating continua with that of electromagnetic and acoustic wave propagation in non-rotating anisotropic chiral media. As mechanical examples a model of a rotating shaft with two degrees of freedom and a continuous model of a rotating circular string passing through the eyelet are studied in detail.
\end{abstract}

\section{Introduction}

\begin{figure}
\includegraphics[width=0.9\textwidth]{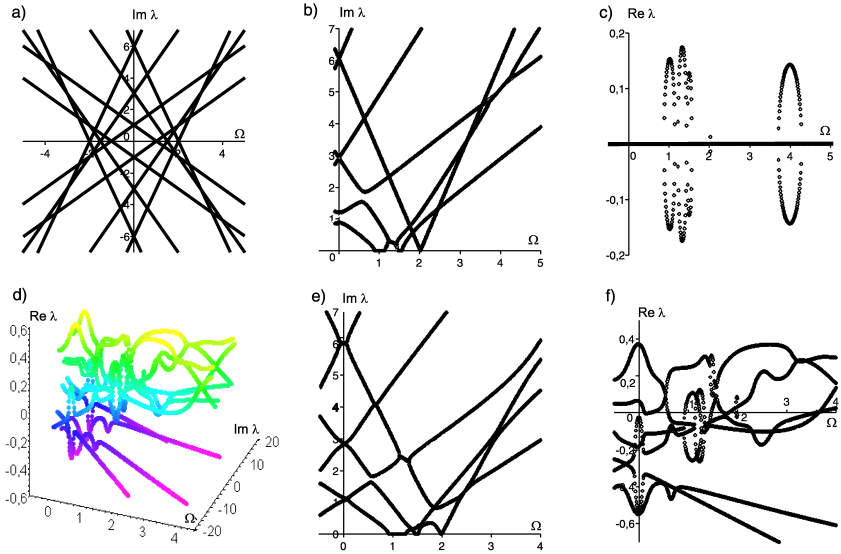}
\caption{\label{fig1} (a) The Campbell diagram of the unperturbed system \rf{i2} with 6 d.o.f. in case of $\omega_1=1$, $\omega_2=3$, and $\omega_3=6$; (b) the Campbell diagram and (c) decay rate plots for the stiffness modification $\kappa{\bf K}_1$ with $\kappa=0.2$; (d) unfolding the Campbell diagram due to perturbation with the matrices ${\bf K}={\bf K}_1$, ${\bf D}={\bf D}_1$, ${\bf N}={\bf N}_1$ and $\kappa=0.2$, $\delta=0.1$, and $\nu=0.2$, (e) the corresponding Campbell diagram and (f) decay rate plots.}
\end{figure}

Bending waves propagate in the circumferential direction of an elastic body of revolution rotating about its axis of symmetry \cite{B1890,LS21,SL71,G07}. The frequencies of the waves plotted against the rotational speed are referred to as {\it the Campbell diagram} \cite{C24,G07}. The spectrum of a perfect rotationally symmetric rotor at standstill has infinitely many double semi-simple eigenvalues---{\it the doublet modes}.  Indeed, for $\mathbb{R}^{2\times 2} \ni {\bf A}={\rm diag}(\omega_1^2,\omega_2^2)$ and ${\bf R}=\left(\begin{array}{cc}
\cos\theta & \sin\theta \\
-\sin\theta & \cos\theta \\
\end{array}
\right)
$
the restriction ${\bf R}^T {\bf A} {\bf R}={\bf A}$ imposed by equivariance of the equations of motion with respect to the action of the circle group implies $\omega_1^2=\omega_2^2$, see, e.g., \cite{DMM92,NN98}.
By this reason the Campbell diagram contains the eigenvalue branches originated after splitting of the doublets by gyroscopic forces \cite{B1890}. The branches correspond to simple pure imaginary eigenvalues and intersect each other forming a {\it spectral mesh} \cite{GK06} in the frequency-speed plane with the doublets at the nodes, Fig.~\ref{fig1}(a).
Perturbations of the axially symmetric rotor by dissipative, conservative, and non-conservative positional forces, caused by its contact with the anisotropic stator, generically unfold the spectral mesh of pure imaginary eigenvalues of the Campbell diagram into separate branches of complex eigenvalues in the $(\Omega,{\rm Im}\lambda,{\rm Re}\lambda)$-space, see Fig.~\ref{fig1}(d).
Nevertheless, the eigenvalue branches in the perturbed Campbell diagram can both avoid crossings and cross each other, Fig.~\ref{fig1}(e). Moreover, the real parts of the perturbed eigenvalues plotted against the rotational speed---{\it decay rate plots} \cite{G07}---can also intersect each other and inflate into `bubbles', Fig.~\ref{fig1}(f). This complicated behavior is difficult to predict and even to interpret according to the studies of numerous mechanical systems \cite{OCB91,CB92,MC95,YH95,TH99,XCY02,CP06,YL06,G07,LST07,SHKH09}. The present work reveals that the unfolding of the Campbell diagrams is determined by a limited number of local
scenarios for eigenvalues as a function of parameters, which form stratified manifolds.

\section{A model of a weakly anisotropic rotor system}

In general, the imperfections in the rotor and stator complicate the linearized equations of motion making them non-self-adjoint with time-dependent coefficients.
Nevertheless, an axially symmetric rotor with an anisotropic stator as well as an
asymmetric rotor with an isotropic stator are autonomous non-conservative gyroscopic systems \cite{G07}.
Neglecting the centrifugal stiffness without loss of generality, we consider the finite-dimensional anisotropic rotor system
\be{i1}
\ddot{\bf x} + (2\Omega{\bf G}+\delta{\bf D})\dot{\bf x} + ({\bf P}+\Omega^2{\bf G}^2+\kappa{\bf K}+\nu{\bf N}) {\bf x}=0,
\ee
which is a perturbation of the isotropic one \cite{NN98}
\be{i2}
\ddot{\bf x} + 2\Omega{\bf G}\dot{\bf x} + ({\bf P}+\Omega^2{\bf G}^2) {\bf x}=0,
\ee
where ${\bf x}=\mathbb{R}^{2n}$, ${\bf P}$=${\rm diag}(\omega_1^2,\omega_1^2,\omega_2^2,\omega_2^2,\ldots,\omega_n^2,\omega_n^2)$ is the stiffness matrix, and ${\bf G}=-{\bf G}^T$ is the matrix of gyroscopic forces defined as
\be{i3}
{\bf G}={\rm blockdiag}({\bf J},2{\bf J},\ldots,n{\bf J}),\quad
{\bf J}=\left(
\begin{array}{rr}
                                                                                        0 & -1 \\
                                                                                        1 & 0 \\
                                                                                      \end{array}
                                                                                    \right).
\ee
The matrices of non-Hamiltonian perturbation corresponding to velocity-dependent dissipative forces, ${\bf D}={\bf D}^T$,
and non-conservative positional forces, ${\bf N}=-{\bf N}^T$, as well as the matrix ${\bf K}={\bf K}^T$ of the Hamiltonian perturbation that breaks the rotational symmetry, can depend on the rotational speed $\Omega$.
The intensity of the perturbation is controlled by the parameters $\delta$, $\kappa$, and $\nu$. Putting $\kappa=0$ and $\nu=0$ in \rf{i1} yields the model considered in \cite{NN98}.

At $\Omega=0$ the eigenvalues $\pm i\omega_s$, $\omega_s>0$, of the isotropic rotor \rf{i2} are double semi-simple with two linearly independent eigenvectors. The sequence of the frequencies $\omega_s$, where $s$ is an integer index, is usually different for various bodies of revolution.
For example, $\omega_s=s$ corresponds to the natural frequency $f_s=\frac{s}{2\pi r}\sqrt{\frac{P}{\rho}}$ of a circular string
of radius $r$, circumferential tension $P$, and mass density $\rho$ per unit length \cite{YH95}.

Substituting ${\bf x}={\bf u}\exp(\lambda t)$ into \rf{i2}, we arrive at the eigenvalue problem
\be{i4}
{\bf L}_0(\Omega){\bf u}:=({\bf I}\lambda^2 + 2\Omega{\bf G}\lambda + {\bf P}+\Omega^2{\bf G}^2) {\bf u}=0.
\ee
The eigenvalues of the operator ${\bf L}_0$ are found in the explicit form
\be{i5}
\lambda_s^{+}= i \omega_s + i s\Omega, \quad \overline{\lambda_s^{-}}= -i \omega_s + i s\Omega,\quad
\lambda_s^{-}= i \omega_s - i s\Omega, \quad
\overline{\lambda_s^{+}}= -i \omega_s - i s\Omega,
\ee
where the overbar denotes complex conjugate. The eigenvectors of $\lambda_s^{+}$ and $\overline{\lambda_s^{-}}$ are
\be{i7}
{\bf u}_1^+=(-i,1,0,0,\ldots,0,0)^T, ~~ \ldots,~~  {\bf u}_n^+=(0,0,\ldots,0,0,-i,1)^T,
\ee
where the imaginary unit holds the $(2s-1)$st position in the vector ${\bf u}_s^+$.
The eigenvectors, corresponding to the eigenvalues $\lambda_s^{-}$ and $\overline{\lambda_s^{+}}$, are simply ${\bf u}^-_s=\overline{{\bf u}^+_s}$

For $\Omega>0$, simple eigenvalues $\lambda_s^{+}$ and $\lambda_s^{-}$ correspond to the forward and backward traveling waves, respectively,
that propagate in the circumferential direction of the rotor.
At the angular velocity
$
\Omega_s^{cr}={\omega_s}/{s}
$
the frequency of the $s$th backward traveling wave vanishes to zero, so that the wave remains stationary in the non-rotating
frame.  We assume further in the text that the sequence of the doublets $i\omega_s$ has the property
$\omega_{s+1}-\omega_s\ge\Omega_s^{cr}$,
which implies the existence of the minimal \textit{critical} speed $\Omega_{cr}=\Omega_1^{cr}=\omega_1$. When the speed of rotation exceeds the critical speed, some backward waves, corresponding to
the eigenvalues $\overline{\lambda_s^{-}}$,
travel slower than the disc rotation speed and appear to be traveling forward (reflected waves).

In Fig.~\ref{fig1}(a) the spectral mesh \rf{i5} is shown for the 6 d.o.f.-system \rf{i2} with the frequencies $\omega_1=1$, $\omega_2=3$, and $\omega_3=6$ that imitate the distribution of the doublets of a circular ring \cite{CP06}.
To illustrate typical unfolding of the Campbell diagram, we plot in Fig.~\ref{fig1}(d)-(f) the eigenvalues of the 6 d.o.f.-system \rf{i1} with $\kappa=0.2$, $\delta=0.1$, $\nu=0.2$, $\omega_1=1$, $\omega_2=3$, and $\omega_3=6$ for the specific symmetry-breaking matrix ${\bf K}={\bf K}_1$, $({\bf K}_1={\bf K}_1^T)$, whose non-zero entries are $k_{11}=1$, $k_{12}=2$, $k_{13}=1$, $k_{14}=2$, $k_{22}=1$, $k_{23}=3$, $k_{24}=4$, $k_{33}=-3$, $k_{44}=-2.5$, $k_{55}=4$, $k_{66}=2$, and for the matrices ${\bf D} =\bf{D}_1$ and ${\bf N} =\bf{N}_1$, where
\be{i9}
{\bf D}_1=\left(
          \begin{array}{rrrrrr}
            -1 & 2 & 1 & 7 & 2 & -2 \\
            2 & 3 & -2 & -4 & 3 & 1 \\
            1 & -2 & 1 & 8 & 2 & 1 \\
            7 & -4 & 8 & 3 & -2 & 3 \\
            2 & 3 & 2 & -2 & 5 & 5 \\
            -2 & 1 & 1 & 3 & 5 & 6 \\
          \end{array}
        \right),~~
{\bf N}_1=\left(
          \begin{array}{rrrrrr}
            0 & -1 & 1 & -1 & -3 & 8 \\
            1 & 0 & 2 & 3 & 2 & 4 \\
            -1 & -2 & 0 & 7 & 1 & 3 \\
            1 & -3 & -7 & 0 & 8 & 2 \\
            3 & -2 & -1 & -8 & 0 & 2 \\
            -8 & -4 & -3 & -2 & -2 & 0 \\
          \end{array}
        \right).\nn
\ee

In the following we classify and interpret the typical behavior of the eigenvalues of weakly anisotropic rotor system \rf{i1} with the use of the perturbation formula for the doublets of the spectral mesh \rf{i5}, which we derive in the next section.

\section{Perturbation of the doublets}

Introducing the indices $\alpha,\beta,\varepsilon,\sigma=\pm 1$ we find that the eigenvalue branches $\lambda_s^{\varepsilon}= i\alpha \omega_s + i \varepsilon s\Omega$ and $\lambda_t^{\sigma}= i\beta \omega_t + i \sigma t\Omega$ cross each other at $\Omega=\Omega_0$ with the origination of the double eigenvalue $\lambda_0=i\omega_0$ with two linearly-independent eigenvectors ${\bf u}_s^{\varepsilon}$
and ${\bf u}_t^{\sigma}$, where
\be{p5}
\Omega_0=\frac{\alpha \omega_s -\beta \omega_t}{\sigma t - \varepsilon s}, \quad
\omega_0=\frac{\alpha \sigma\omega_s t -\beta \varepsilon \omega_t s}{\sigma t - \varepsilon s}.
\ee

Let ${\bf M}$ be one of the matrices ${\bf D}$, ${\bf K}$, or ${\bf N}$.
In the following, we decompose the matrix ${\bf M}\in \mathbb{R}^{2n\times2n}$ into $n^2$ blocks ${\bf M}_{st}\in \mathbb{R}^{2\times2}$, where
$s,t=1,2,\ldots,n$
\be{p6}
{\bf M}=\left(
          \begin{array}{ccccc}
            * & * & * & * & *\\
            * & {\bf M}_{ss} & \cdots & {\bf M}_{st}& *\\
            * & \vdots & \ddots & \vdots& *\\
            * & {\bf M}_{ts} & \cdots & {\bf M}_{tt}& *\\
            * & * & * & * & *\\
          \end{array}
        \right),\quad {\bf M}_{st}=\left(
               \begin{array}{ll}
                 m_{2s-1,2t-1} & m_{2s-1,2t} \\
                 m_{2s,2t-1} & m_{2s,2t} \\
               \end{array}
             \right).
\ee
Note that ${\bf D}_{st}={\bf D}_{ts}^T$, ${\bf K}_{st}={\bf K}_{ts}^T$, and ${\bf N}_{st}=-{\bf N}_{ts}^T$.

We consider a general perturbation of the matrix operator of the isotropic rotor ${\bf L}_0(\Omega)+\Delta {\bf L}(\Omega)$. The size of the perturbation $\Delta {\bf
L}(\Omega)=\delta\lambda{\bf D}+\kappa{\bf K}+\nu {\bf N}\sim
\epsilon$ is small, where $\epsilon=\| \Delta{\bf L}(\Omega_0) \|$ is
the Frobenius norm of the perturbation at $\Omega=\Omega_0$.
For small $\Delta\Omega=|\Omega-\Omega_0|$ and $\epsilon$ the increment to the doublet
$\lambda_0=i\omega_0$ with the eigenvectors ${\bf u}_s^{\varepsilon}$ and ${\bf u}_t^{\sigma}$, is given by the formula
$
\det({\bf R}+(\lambda-\lambda_0){\bf Q})=0
$
\cite{KMS05a,Ki08}, where the entries of the  $2\times2$ matrices $\bf Q$ and $\bf R$ are
\ba{p8}
Q_{st}^{\varepsilon \sigma}&{=}&2i\omega_0(\bar{\bf u}_s^{\varepsilon})^T{\bf u}_t^{\sigma}{+}2\Omega_0(\bar{\bf u}_s^{\varepsilon})^T{\bf G}{\bf u}_t^{\sigma}, \nn \\
R_{st}^{\varepsilon \sigma}&{=}&(2i\omega_0 (\bar{\bf u}_s^{\varepsilon})^T{\bf G}{\bf u}_t^{\sigma}{+}2\Omega_0 (\bar{\bf u}_s^{\varepsilon})^T{\bf G}^2{\bf u}_t^{\sigma})(\Omega{-}\Omega_0)\nn\\&{+}& i\omega_0(\bar{\bf u}_s^{\varepsilon})^T{\bf D}{\bf u}_t^{\sigma}\delta{+} (\bar{\bf u}_s^{\varepsilon})^T{\bf K}{\bf u}_t^{\sigma}\kappa{+} (\bar{\bf u}_s^{\varepsilon})^T{\bf N}{\bf u}_t^{\sigma}\nu.
\ea
Calculating the coefficients \rf{p8} with the eigenvectors \rf{i7} we find the real and imaginary parts of the sensitivity of the doublet $\lambda_0=i\omega_0$ at the crossing \rf{p5}
\ba{p12}
{\rm Re}\lambda&=&-\frac{1}{8}\left(\frac{{\rm Im}A_1}{\alpha \omega_s}+\frac{{\rm Im}B_1}{\beta \omega_t} \right) \pm
\sqrt{\frac{|c|-{\rm Re}c}{2}},\nn\\
{\rm Im}\lambda&=&\omega_0+\frac{\Delta\Omega}{2}(s\varepsilon + t\sigma)+\frac{\kappa}{8}\left(\frac{{\rm tr}{\bf K}_{ss}}{\alpha \omega_s}+\frac{{\rm tr}{\bf K}_{tt}}{\beta \omega_t} \right)\pm
\sqrt{\frac{|c|+{\rm Re}c}{2}},
\ea
where $c={\rm Re}c+i{\rm Im}c$ with
\ba{p12a}
{\rm Im}c &=& \frac{\alpha\omega_t {\rm Im}A_1 -\beta \omega_s{\rm Im}B_1 }{8\omega_s\omega_t}(s\varepsilon-t\sigma)\Delta\Omega\nn\\&+&
\kappa\frac{(\alpha\omega_s{\rm tr}{\bf K}_{tt}-\beta\omega_t{\rm tr}{\bf K}_{ss})(\alpha\omega_s {\rm Im}B_1-\beta\omega_t{\rm Im}A_1)}{32\omega_s^2\omega_t^2}\nn\\
&-&\alpha\beta\kappa\frac{{\rm Re}A_2 {\rm tr}{\bf K}_{st}{\bf J}_{\varepsilon \sigma}-{\rm Re}B_2 {\rm tr}{\bf K}_{st}{\bf I}_{\varepsilon \sigma}}
{8\omega_s\omega_t},\nn \\
{\rm Re}c &=& \left(\frac{t\sigma-s\epsilon}{2}\Delta\Omega+\kappa\frac{\beta\omega_s{\rm tr}{\bf K}_{tt}-
\alpha\omega_t{\rm tr}{\bf K}_{ss}}{8\omega_s\omega_t}\right)^2\nn\\&+&\alpha\beta\frac{( {\rm tr}{\bf K}_{st}{\bf J}_{\varepsilon \sigma})^2+( {\rm tr}{\bf K}_{st}{\bf I}_{\varepsilon \sigma})^2}{16\omega_s\omega_t}\kappa^2\nn \\
&-&\frac{(\alpha\omega_s {\rm Im}B_1-\beta\omega_t{\rm Im}A_1)^2+4\alpha\beta\omega_s\omega_t (({\rm Re}A_2)^2+({\rm Re}B_2)^2)}
{64\omega_s^2\omega_t^2}.
\ea
The coefficients $A_{1}$, $A_{2}$ and $B_{1}$, $B_{2}$ depend only on those entries of the matrices $\bf D$, $\bf K$, and $\bf N$ that belong to the four $2\times2$ blocks \rf{p6} with the indices $s$ and $t$
\ba{p13}
A_1&=&\delta \lambda_0 {\rm tr}{\bf D}_{ss}{+}\kappa{\rm tr}{\bf K}_{ss}{+}\varepsilon 2 i \nu n_{2s-1,2s},\nn\\
A_2&=&\sigma\nu  {\rm tr}{\bf N}_{st}{\bf I}_{\varepsilon \sigma} {+}i(\delta \lambda_0 {\rm tr}{\bf D}_{st}{\bf J}_{\varepsilon \sigma}{+}\kappa {\rm tr}{\bf K}_{st}{\bf J}_{\varepsilon \sigma}),\nn\\
B_1&=&\delta \lambda_0 {\rm tr}{\bf D}_{tt}{+}\kappa{\rm tr}{\bf K}_{tt}{+}\sigma 2 i \nu n_{2t-1,2t},\nn\\
B_2&=&\sigma\nu {\rm tr}{\bf N}_{st}{\bf J}_{\varepsilon \sigma} {-}i(\delta \lambda_0  {\rm tr}{\bf D}_{st}{\bf I}_{\varepsilon \sigma}{+}\kappa {\rm tr}{\bf K}_{st}{\bf I}_{\varepsilon \sigma}),
\ea
where
\be{p10}
{\bf I}_{\varepsilon \sigma}=\left(
                               \begin{array}{cc}
                                 \varepsilon & ~~0 \\
                                 0 & ~~\sigma \\
                               \end{array}
                             \right),\quad
{\bf J}_{\varepsilon \sigma}=\left(
                               \begin{array}{cc}
                                 0 & ~-\sigma \\
                                 \varepsilon & ~~~0 \\
                               \end{array}
                             \right).
\ee
Therefore, we have identified the elements of the perturbing matrices that control practically important {\it eigenvalue assignment} \cite{O08} near every particular node  $(\Omega_0,\omega_0)$ of the spectral mesh.

\section{MacKay's eigenvalue cones and instability bubbles}

Modification of the stiffness matrix induced by the elastic support interacting with the rotating continua is typical in the models of rotating shafts \cite{SM68}, computer disc drives \cite{OCB91,CB92}, circular saws \cite{YH95,TH99,XCY02}, car brakes \cite{MC95,HS08,SHKH09}, and turbine wheels \cite{G07,LST07}.

\begin{figure}
\includegraphics[width=0.99\textwidth]{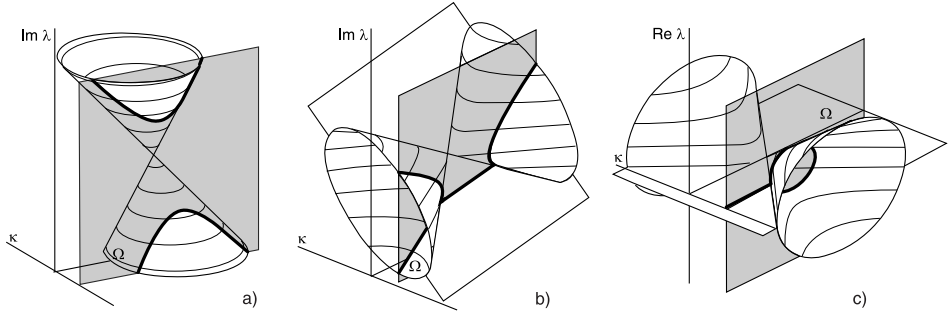}
\caption{Eigenvalue surfaces (MacKay, 1986) and (bold lines) their cross-sections in the plane $\kappa=const$ (grey):
(a) a near-vertically oriented cone ${\rm Im}\lambda(\Omega, \kappa)$ in the subcritical range (${\rm Re}\lambda=0$);
(b) imaginary parts forming a near-horizontally oriented cone \rf{c1} with the attached membrane \rf{c3} and
(c) the real parts forming a near-horizontally oriented cone $({\rm Re}\lambda)^2=-{\rm Re}c$ with the attached membrane
${\rm Re}\lambda=0$ in the supercritical range.}
\label{fig2}
\end{figure}

Assuming $\delta=0$ and $\nu=0$ in \rf{p12} we find that  the eigenvalues of the system \rf{i5} with the stiffness modification $\kappa{\bf K}$ either are pure imaginary $({\rm Re}\lambda=0)$ and form a conical surface in the $(\Omega,\kappa,{\rm Im}\lambda)$-space with the apex at the point $(\Omega_0,0,\omega_0)$
\be{c1}
\left({\rm Im}\lambda-\omega_0 - \frac{\kappa}{8} \left(\frac{{\rm tr}{\bf K}_{ss}}{\alpha\omega_s}+\frac{{\rm tr}{\bf K}_{tt}}{\beta\omega_t}\right)-\frac{\Omega-\Omega_0}{2}(s\varepsilon + t \sigma) \right)^2={\rm Re}c,
\ee
see Fig.~\ref{fig2}(a), or they are complex and in the $(\Omega,\kappa,{\rm Re}\lambda)$-space their real parts originate a cone $({\rm Re}\lambda)^2=-{\rm Re}c$ with the apex at the point $(\Omega_0,0,0)$, Fig.~\ref{fig2}(c). In the $(\Omega,\kappa,{\rm Im}\lambda)$-space the corresponding imaginary parts belong to the plane
\be{c3}
{\rm Im}\lambda=\omega_0+\frac{\kappa}{8} \left(\frac{{\rm tr}{\bf K}_{ss}}{\alpha\omega_s}+\frac{{\rm tr}{\bf K}_{tt}}{\beta\omega_t}\right)+\frac{\Omega-\Omega_0}{2}(s\varepsilon + t \sigma),
\ee
which is attached to the cone \rf{c1} as shown in Fig.~\ref{fig2}(b).

The existence of eigenvalues with ${\rm Re}\lambda \ne 0$ depends on the sign of  $\alpha\beta$.
It is negative only if the crossing in the Campbell diagram is formed by the eigenvalue branch of the reflected wave
and by that of either forward- or backward traveling wave. Otherwise, $\alpha\beta>0$.
Due to the property $\omega_{s+1}-\omega_s\ge\Omega_s^{cr}$ the crossings of the reflected wave with the forward- and backward traveling waves occur only
in the {\it supercritical} speed range $|\Omega|\ge\Omega_{cr}$.
The crossings with $\alpha\beta>0$ are situated in both the super- and {\it subcritical} $(|\Omega|<\Omega_{cr})$ ranges.
Therefore, the eigenvalues with ${\rm Re}\lambda \ne 0$ originate only near the supercritical crossings of the eigenvalue branches $\lambda_s^{\varepsilon}$ and $\lambda_t^{\sigma}$ with $\alpha\beta<0$, when the parameters in the $(\Omega,\kappa)$-plane are in the sector ${\rm Re}c<0$ bounded by the straight lines ${\rm Re}c=0$
\be{c5}
\kappa=\frac{4(s\varepsilon-t\sigma)(\Omega-\Omega_0)}{ \frac{k_{2t-1,2t-1}{+}k_{2t,2t}}{\beta\omega_t}{-}\frac{k_{2s-1,2s-1}{+}k_{2s,2s}}{\alpha\omega_s}{\pm}2\sqrt{\frac{(\varepsilon k_{2s-1,2t-1}{+}\sigma k_{2s,2t})^2+(\varepsilon k_{2s-1,2t}{-}\sigma k_{2s,2t-1})^2}{ -\alpha\beta\omega_s\omega_t}}}.
\ee

Since for $\alpha\beta<0$ the cones of the real parts $({\rm Re}\lambda)^2=-{\rm Re}c$ are near-horizontally oriented and extended along the $\kappa$-axis in the $(\Omega,\kappa,{\rm Re}\lambda)$-space, their cross-sections by the planes $\kappa=const$ are ellipses, as shown in Fig.~\ref{fig1}(c) and in Fig.~\ref{fig2}(c). Since a part of the ellipse corresponds to the eigenvalues with positive real parts, the ellipse is called the \textit{bubble of instability} \cite{MS86}. Equation \rf{c5} is, therefore, a linear approximation to the boundary of the domain of instability, which is divergence (parametric resonance) for $\Omega_0=\Omega_s^{cr}$ and flutter (combination resonance) otherwise. The near-horizontal orientation of the corresponding cones of imaginary parts \rf{c1} in the $(\Omega,\kappa,{\rm Im}\lambda)$-space explains deformation in the presence of the perturbation $\kappa{\bf K}$ of the crossings with $\alpha\beta<0$ into the branches of a hyperbola connected by a straight line in the Campbell diagram, see Fig.~\ref{fig1}(b) and Fig.~\ref{fig2}(b).

Near the crossings with $\alpha\beta>0$ the perturbed eigenvalues are pure imaginary (stability). The corresponding cones of imaginary parts \rf{c1} are near-vertically oriented in the $(\Omega,\kappa,{\rm Im}\lambda)$-space, Fig.~\ref{fig2}(a). In the plane $\kappa=const$ this yields the \textit{avoided crossing} \cite{MK86,MS86}, which is approximated by a hyperbola shown by the bold lines in Fig.~\ref{fig2}(a) (cf. Fig.~\ref{fig1}(b)).

The conical singularities of the eigenvalue surfaces in the Hamiltonian systems are traced back to the works of Hamilton himself, who predicted the effect of conical refraction of light in birefringent crystals \cite{H1833,BJ07}. Later on, the conical singularities of eigenvalue surfaces were found in atomic, nuclear, and molecular physics \cite{NW29,T37,MH93}. Nowadays they bear a name of the Hamilton's {\it diabolical points} \cite{BJ07}. The existence of the two different orientations of the eigenvalue cones in the Hamiltonian systems was established in \cite{MK86}. This result is based on the works of \citeasnoun{W36} and \citeasnoun{K83}, who introduced the signature of eigenvalues known as the {\it symplectic signature} in the Hamiltonian mechanics \cite{MS98} and as the {\it Krein signature} in a broader context of the theory of Krein spaces \cite{KGS09}.

To evaluate the symplectic signatures, we reduce \rf{i2} to $\dot{\bf y}={\bf A}{\bf y}$, where
\be{c6}
{\bf A}=\left(
                                           \begin{array}{rr}
                                             -\Omega {\bf G} & {\bf I}_n \\
                                             -{\bf P} & -\Omega {\bf G} \\
                                           \end{array}
                                         \right)={\bf J}_{2n}{\bf A}^T{\bf J}_{2n},~~
                                         {\bf J}_{2n}=\left(
                                                   \begin{array}{cc}
                                                     0 & -{\bf I}_n \\
                                                     {\bf I}_n & 0 \\
                                                   \end{array}
                                                 \right),~~
                                         {\bf y}=\left(
          \begin{array}{c}
            {\bf x} \\
            \dot{\bf x}+\Omega{\bf G}{\bf x} \\
          \end{array}
        \right).
\ee
The Hamiltonian symmetry of the matrix $\bf A$
implies its self-adjointness in a Krein space with the indefinite inner product
$
[{\bf a},{\bf b}]=\overline{\bf b}^T {\bf J}_{2n}{\bf a}, \quad {\bf a},{\bf b}\in \mathbb{C}^{2n}.
$
The matrix $\bf A$ has the eigenvalues  $\lambda_s^{\pm}$ given by the formulas \rf{i5} with the eigenvectors
\be{c8}
{\bf a}_s^{++}=\left(
      \begin{array}{c}
        {\bf u}_s^+ \\
        \lambda_s^+ {\bf u}_s^++\Omega{\bf G}{\bf u}_s^+ \\
      \end{array}
    \right),\quad
{\bf a}_s^{+-}=\left(
      \begin{array}{c}
        {\bf u}_s^- \\
        \lambda_s^- {\bf u}_s^-+\Omega{\bf G}{\bf u}_s^- \\
      \end{array}
    \right),
\ee
where the vectors ${\bf u}_s^{\pm}$ are determined by expressions \rf{i7}. Since
$
i[{\bf a}_s^{++},{\bf a}_s^{++}]=i[{\bf a}_s^{+-},{\bf a}_s^{+-}]=4\omega_s>0,
$
the eigenvalues $\lambda_s^+$ and $\lambda_s^-$ of the forward and backward traveling waves
acquire  \textit{positive symplectic (Krein) signature}. The eigenvalues $\overline{\lambda_s^+}$ and $\overline{\lambda_s^-}$ of the reflected waves with
$
i[{\bf a}_s^{-+},{\bf a}_s^{-+}]=i[{\bf a}_s^{--},{\bf a}_s^{--}]=-4\omega_s<0,
$
have the opposite, \textit{negative symplectic (Krein) signature} \cite{MK86,MS98}. The signature of an eigenvalue in the Campbell diagram coincides with the sign of the doublet at $\Omega=0$, from which it is branched, and does not change with the variation of $\Omega$. This implies $\alpha\beta>0$ and near-vertically oriented cones of imaginary parts \rf{c1} at the crossings of eigenvalue branches with \textit{definite} (positive) signature and $\alpha\beta<0$ and near-horizontally oriented cones of imaginary parts \rf{c1} at the crossings with \textit{mixed} signature \cite{MK86}.

The symplectic signature coincides with the sign of the second derivative of the energy, which is a non-degenerate definite  quadratic form on the
real invariant space associated to a complex conjugate pair of simple pure imaginary non-zero
eigenvalues \cite{MK86}.  Interaction of waves with positive and negative energy is a well known mechanism of instability of the moving fluids and plasmas \cite{MK86,SF89,HF08}; in rotor dynamics this yields flutter in the supercritical speed range, which is known as the mass and stiffness instabilities \cite{MC95,G07}.

Therefore, in case when anisotropy of the stator is caused by the stiffness modification only, the unfolding of the Campbell diagram is completely described by one-parameter slices of the two-parameter MacKay's eigenvalue cones. Since there are only two possible spatial orientations of the cones corresponding to either definite or mixed symplectic signatures, all one has to do to predict the unfolding of the Campbell diagram into avoided crossings or into bubbles of instability is to calculate the signatures of the appropriate eigenvalues of the isotropic rotor. In the following, we develop the MacKay's theory further and show that even in the presence of non-Hamiltonian perturbations, all the observed peculiarities of the Campbell diagrams and decay rate plots are one-parameter slices of the eigenvalue surfaces near a limited number of other singularities whose type is dictated by the definiteness of the symplectic signature of the double eigenvalues at the crossings.

\section{Double coffee filter singularity near the crossings with definite symplectic (Krein) signature}

Understanding general rules of unfolding the Campbell diagrams of weakly anisotropic rotor systems in the presence of dissipative and non-conservative perturbations is important for linear stability analysis and for interpretation of numerical data in both low- and high-speed applications \cite{G07}. In the latter \textit{supercritical flutter and divergence} instabilities are easily excited near the crossings with the mixed symplectic signature just by the Hamiltonian perturbations like stiffness modification. In  low-speed applications unfolding of the Campbell diagram is directly related to the  onset of friction-induced oscillations in brakes, clutches, paper calenders, and even in musical instruments like the glass harmonica \cite{Sp61,OM01,Ki08,KKS08,S08,HS08,O08,SHKH09}. In contrast to the supercritical instabilities, excitation of the \textit{subcritical flutter} near the crossings with the definite symplectic signature by the Hamiltonian perturbations only, is impossible. In this case the non-Hamiltonian dissipative and circulatory forces are required for destabilization.

In general, dissipative, $\delta{\bf D}$, and non-conservative positional, $\nu{\bf N}$, perturbations
unfold the MacKay's eigenvalue cones \rf{c1} and $({\rm Re}\lambda)^2=-{\rm Re}c$ into the surfaces ${\rm Im}\lambda(\Omega,\kappa)$ and ${\rm Re}\lambda(\Omega,\kappa)$,
described by formulas \rf{p12}. The new eigenvalue surfaces have singularities at the {\it exceptional points} \cite{KKM03,BD03}. The latter correspond to the double eigenvalues with the Jordan chain that born from the parent semi-simple doublet $i\omega_0$ at $\Omega=\Omega_0$.
In some works numerical methods were developed to find the coordinates of these singularities \cite{J88,S07}.
Perturbation of the Hamilton's diabolical points is another efficient way to locate exceptional points \cite{KMS05a}. Indeed, condition $c=0$ yields their approximate loci in the $(\Omega,\kappa)$-plane
\be{cf1}
\Omega_{EP}^{\pm}=\Omega_0\pm\frac{4\omega_s\omega_tU-\beta\omega_s{\rm tr}{\bf K}_{tt}+\alpha\omega_t{\rm tr}{\bf K}_{ss}}{4\omega_s\omega_t(t\sigma-s\varepsilon)}\sqrt{\frac{N}{D}},~~
\kappa_{EP}^{\pm}=\pm\sqrt{\frac{N}{D}},
\ee
where
\ba{cf2}
U&{=}&\frac{{\rm Re}A_2 {\rm tr}{\bf K}_{st}{\bf J}_{\varepsilon \sigma}-{\rm Re}B_2 {\rm tr}{\bf K}_{st}{\bf I}_{\varepsilon \sigma}}{\alpha\omega_s {\rm Im}B_1-\beta\omega_t{\rm Im}A_1},\\
D &{=}&U^2+\alpha\beta\left[\left(\frac{{\rm tr}{\bf K}_{st}{\bf J}_{\varepsilon \sigma}}{2\sqrt{\omega_s\omega_t}}\right)^2 {+} \left(\frac{{\rm tr}{\bf K}_{st}{\bf I}_{\varepsilon \sigma}}{2\sqrt{\omega_s\omega_t}}\right)^2\right],\nn\\
N &{=}&\left(\frac{\alpha\omega_s {\rm Im}B_1-\beta\omega_t{\rm Im}A_1}
{4\omega_s\omega_t}\right)^2 +\alpha\beta\left[\left(\frac{{\rm Re}A_2}{2\sqrt{\omega_s\omega_t}}\right)^2 {+}\left(\frac{{\rm Re}B_2}{2\sqrt{\omega_s\omega_t}}\right)^2\right].\nn
\ea
The crossings with the definite symplectic signature $(\alpha\beta>0)$ always produce a pair of the exceptional points.  For example, for pure non-conservative $(\delta=0)$ and pure dissipative $(\nu=0)$ perturbation of the doublets at $\Omega_0=0$, formulas \rf{cf1} read
\ba{cf3}
\Omega_{EP,n}^{\pm}&=&0,\quad \kappa_{EP,n}^{\pm}=\pm\frac{2\nu n_{2s-1,2s}}{\rho_1({\bf K}_{ss})-\rho_2({\bf K}_{ss})};\nn\\
\Omega_{EP,d}^{\pm}&=&\pm\delta\frac{\mu_1({\bf D}_{ss})-\mu_2({\bf D}_{ss})}{4s},\quad \kappa_{EP,d}^{\pm}=0,
\ea
where $\rho_{1,2}({\bf K}_{ss})$ are the eigenvalues of the block ${\bf K}_{ss}$ of the matrix $\bf K$
and $\mu_{1,2}({\bf D}_{ss})$ are those of the block ${\bf D}_{ss}$ of $\bf D$. In case of the mixed symplectic signature $(\alpha\beta<0)$ the two exceptional points exist when $N/D>0$ and do not exist otherwise.

\begin{figure}
\includegraphics[width=0.99\textwidth]{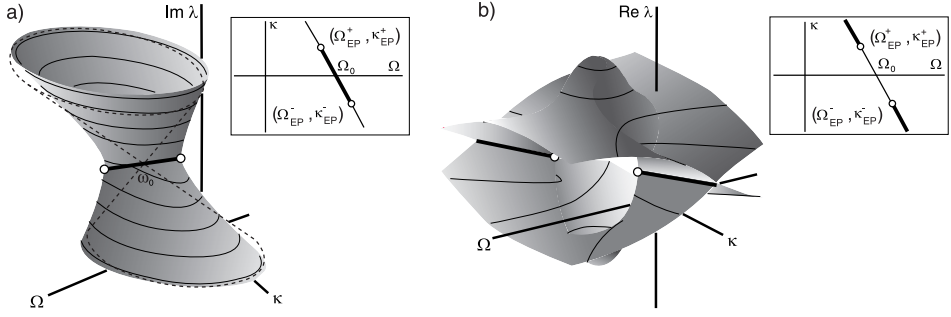}
\caption{\label{fig3} (a) The `double coffee filter' singular surface ${\rm Im}\lambda(\Omega,\kappa)$ with the exceptional points (open circles) and  branch cut (bold lines) originated from the MacKay's cone (dashed lines) due to mixed dissipative and circulatory perturbation at any crossing with the definite symplectic signature; (b) the corresponding `viaduct'  ${\rm Re}\lambda(\Omega,\kappa)$. }
\end{figure}

Strong influence of the exceptional points on the stability and their relation to the Ziegler's destabilization paradox due to small damping is well recognized \cite{B56,Ki07,KM07,K07,S08}.
In numerous applications in rotor dynamics \cite{OCB91,CB92,MC95,YH95,TH99,XCY02,G07,LST07} as well as in hydrodynamics \cite{O91}, crystal optics \cite{BD03}, acoustics \cite{SS00}, and microwave billiards \cite{KKM03},  the generalized crossing scenario in the vicinity of the exceptional points has been observed (visible also in Fig.~\ref{fig1}(e,f)) when at the same values of the parameters the imaginary parts of the eigenvalues cross, whereas the real parts don't and vice versa. In our setting, the conditions for coincidence of imaginary parts of the eigenvalues \rf{p12} are ${\rm Im}c=0$ and ${\rm Re}c\le0$ and that for coincidence of the real parts are ${\rm Im}c=0$ and ${\rm Re}c\ge0$. Both real and imaginary parts of the eigenvalues coincide only at the two exceptional points $(\Omega_{EP}^{+},\kappa_{EP}^{+})$ and $(\Omega_{EP}^{-},\kappa_{EP}^{-})$.
The segment of the line ${\rm Im}c=0$ connecting the exceptional points is the projection of the branch cut of a singular eigenvalue surface ${\rm Im}\lambda(\Omega,\kappa)$. The adjacent parts of the line correspond to the branch cuts of the singular eigenvalue surface
${\rm Re}\lambda(\Omega,\kappa)$.
Since simultaneous intersection of the different segments of the line ${\rm Im}c=0$ in the $(\Omega,\kappa)$-plane is not possible one observes the generalized crossing scenario \cite{KKM03,KMS05a} in the planes $(\Omega,{\rm Im}\lambda)$ and $(\Omega,{\rm Re}\lambda)$ or $(\kappa,{\rm Im}\lambda)$ and $(\kappa,{\rm Re}\lambda)$.

For example, in case of pure non-conservative positional perturbation the real parts of the eigenvalues developing near the doublets  at $\Omega_0=0$ cross each other in the $(\Omega,{\rm Re}\lambda)$-plane at the points of the branch cuts
$\kappa^2>(\kappa_{EP,n}^{\pm})^2$
\be{cf4}
{\rm Re}\lambda=\pm\frac{ 2\nu s n_{2s-1,2s}}{(\rho_1({\bf K}_{ss})-\rho_2({\bf K}_{ss}))\sqrt{\kappa^2-(\kappa_{EP,n}^{\pm}})^2}\Omega+O(\Omega^{3}),
\ee
whereas for $\kappa^2<(\kappa_{EP,n}^{\pm})^2$ they avoid crossing
\be{cf5}
{\rm Re}\lambda=\pm \frac{\rho_1({\bf K}_{ss})-\rho_2({\bf K}_{ss})}{4\omega_s}\sqrt{(\kappa_{EP,n}^{\pm})^2-\kappa^2}+O(\Omega^{2}).
\ee
At the exceptional points $\kappa=\kappa_{EP,n}^{\pm}$ the eigenvalue branches touch each other
\be{cf6}
{\rm Re}\lambda=\pm \frac{1}{2}\sqrt{\frac{2\nu s n_{2s-1,2s}}{\omega_s} \Omega}+O(\Omega^{3/2}).
\ee
The degenerate crossing \rf{cf6} of the real parts has been observed in the model of a rotating circular string  \cite{YH95,Ki08}.

Pure dissipative perturbation of the doublets at $\Omega_0=0$ yields crossings of the real parts at the branch cuts $\Omega^2>(\Omega_{EP,d}^{\pm})^2$ in the $({\rm Re}\lambda,\kappa)$-plane and veering of the imaginary parts
\ba{cf7}
{\rm Im}\lambda&=&\omega_s\pm s\sqrt{\Omega^2-(\Omega_{EP,d}^{\pm})^2}+O(\kappa),\nn\\
{\rm Re}\lambda&=&-\frac{\delta{\rm tr}{\bf D}_{ss}}{4}\pm\frac{\gamma}{16s\omega_s\sqrt{\Omega^2-(\Omega_{EP,d}^{\pm})^2}}\delta\kappa+O(\kappa^3),
\ea
where $\gamma=2{\rm tr}{\bf K}_{ss}{\bf D}_{ss}-{\rm tr}{\bf K}_{ss} {\rm tr}{\bf D}_{ss}$.
At the branch cut $\Omega^2<(\Omega_{EP,d}^{\pm})^2$
the imaginary parts cross and the real parts avoid crossing
\ba{cf8}
{\rm Im}\lambda&=&\omega_s+\frac{{\rm tr}{\bf K}_{ss}}{4\omega_s}\kappa\pm\frac{\gamma}{16s\omega_s\sqrt{(\Omega_{EP,d}^{\pm})^2-\Omega^2}}\delta\kappa+O(\kappa^2),\nn\\
{\rm Re}\lambda&=&-\frac{\delta{\rm tr}{\bf D}_{ss}}{4}\pm s\sqrt{(\Omega_{EP,d}^{\pm})^2-\Omega^2}+O(\kappa^2).
\ea
At  $\Omega=\Omega_{EP,d}^{\pm}$ the crossings of both real and imaginary parts are degenerate
\be{cf9}
{\rm Re}\lambda=-\frac{\delta{\rm tr}{\bf D}_{ss}}{4}\pm\frac{1}{4}\sqrt{-\delta\kappa\frac{\gamma}{\omega_s}}+O(\kappa^{3/2}),\quad
{\rm Im}\lambda=\omega_s\pm\frac{1}{4}\sqrt{-\delta\kappa\frac{\gamma}{\omega_s}}+\frac{{\rm tr}{\bf K}_{ss}}{4\omega_s}\kappa+O(\kappa^{3/2}).
\ee
The evolving eigenvalue branches reconstruct the eigenvalue surfaces shown in Fig.~\ref{fig3}.
In the one-parameter slices of the surfaces the transformation of the eigenvalue branches from the crossing to the avoided crossing due to variation of the parameters $\Omega$ and $\kappa$ occurs after the passage through the exceptional points, where the branches touch each other and the eigenvalue surfaces have Whitney's umbrella singularities. The surface of the imaginary parts shown in Fig.~\ref{fig3}(a) is formed by the two Whitney's umbrellas with the handles (branch cuts) glued when they are oriented toward each other. This singular surface is known in the physical literature on wave propagation in anisotropic media as the \textit{double coffee filter} \cite{KKM03,BD03}.  The \textit{viaduct} singular surface of the real parts results from the gluing of the roofs of two Whitney's umbrellas when their handles are oriented outwards, Fig.~\ref{fig3}(b).
The double coffee filter singularity is a result of the deformation of the MacKay's eigenvalue cone (shown by the dashed lines in Fig.~\ref{fig3}(a)) by dissipative and non-conservative positional perturbations. The perturbations foliate the plane ${\rm Re}\lambda=0$ into the viaduct which has self-intersections along two branch cuts and an ellipse-shaped arch between the two exceptional points, Fig.~\ref{fig3}(b). Both types of singular surfaces appear when non-Hermitian perturbation of Hermitian matrices is considered \cite{K75,KMS05a}.

{\it Therefore, in a weakly non-Hamiltonian system \rf{i1} the fundamental qualitative effect of the splitting of the doublets with the definite symplectic (Krein) signature is the origination of the double coffee filter of the imaginary parts and the viaduct of the real parts. Structural modification of the matrices of dissipative and non-conservative positional forces generically does not change the type of the surfaces, preserving the exceptional points and the branch cuts.}

\begin{figure}
\includegraphics[width=0.9\textwidth]{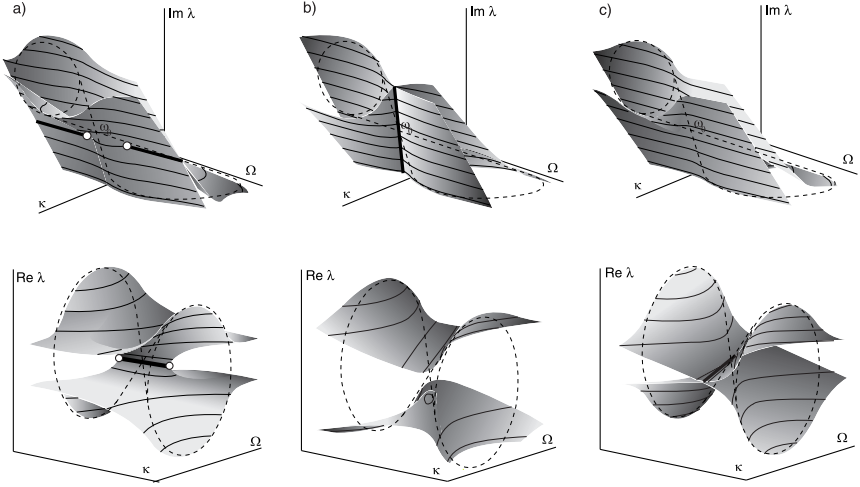}
\caption{\label{fig4a} Mixed symplectic signature $(\alpha\beta<0)$: (a) The viaduct  ${\rm Im}\lambda(\Omega,\kappa)$ and the double coffee filter ${\rm Re}\lambda(\Omega,\kappa)$ for $D<0$ and $N<0$; (b) the surfaces ${\rm Im}\lambda(\Omega,\kappa)$ crossed along the branch cut (bold line) and the separated surfaces ${\rm Re}\lambda(\Omega,\kappa)$ for $D<0$, $N>0$; (c) separated surfaces of imaginary parts and crossed surfaces of real parts for $D>0$, $N<0$.  }
\end{figure}

\section{Unfolding MacKay's cones with mixed signature}

The definite symplectic signature $(\alpha\beta>0)$ implies $D>0$ and $N>0$ and thus uniquely determines the type of the singular surface for the real and imaginary parts of the perturbed eigenvalues. The case of the mixed symplectic signature $(\alpha\beta<0)$ possesses several scenarios for the unfolding of the MacKay's cones by the non-Hamiltonian perturbation, because $D$ and $N$ can have different signs.

When $D>0$ and $N>0$, the imaginary parts of the eigenvalues form the double coffee filter singular surface whereas the real parts originate the viaduct, Fig.~\ref{fig3}. For negative $D$ and negative $N$ the type of the surfaces is interchanged: the imaginary parts form the viaduct and the real parts originate the double coffee filter, Fig.~\ref{fig4a}(a).

Exceptional points are not created for negative values of $N/D$.  In this case the eigenvalue surfaces either intersect each other along the branch cut, which projects into the line ${\rm Im}c=0$ in the $(\Omega,\kappa)$-plane, or do not cross at all. When $N>0$ the surfaces of the imaginary parts ${\rm Im}\lambda(\Omega,\kappa)$ cross and the surface ${\rm Re}\lambda(\Omega,\kappa)$ avoid crossing, Fig.~\ref{fig4a}(b). For $N<0$ the surfaces of the imaginary parts are separated and that of the real parts cross, Fig.~\ref{fig4a}(c).

\section{Example 1. A rotating shaft}

\begin{figure}
\includegraphics[width=0.88\textwidth]{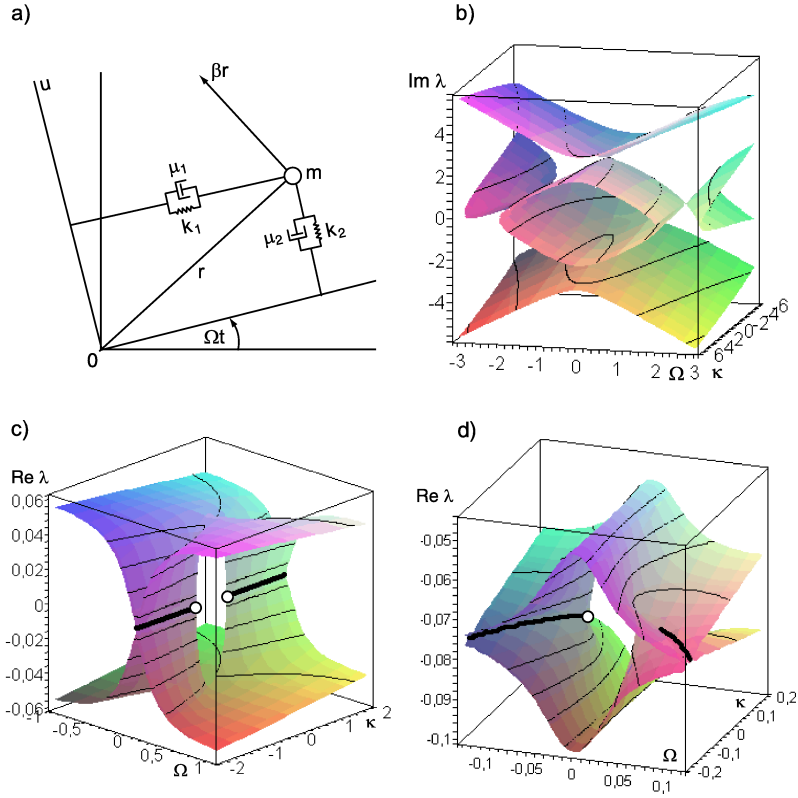}
\caption{\label{fig4} (a) A model of the rotating shaft; (b) four MacKay's cones due to stiffness modification ($\mu_1=0$, $\mu_2=0$, $\beta=0$); (c) the viaduct singular surface created by the circulatory force only ($\beta=0.2$) and (d) by the damping only ($\mu_1=0.1$, $\mu_2=0.2$).  }
\end{figure}

The simplest mechanical systems described by equations \rf{i1} and \rf{i2} are some two-degrees-of-freedom models of rotating shafts \cite{K24,SM68,NN98,G07}.
In \cite{SM68} the shaft is modeled as the mass $m$ which is
attached  by two springs with the stiffness coefficients $k_1$ and $k_2=k_1+\kappa$ and two dampers with the coefficients $\mu_1$ and $\mu_2$ to a coordinate system rotating at constant angular velocity $\Omega$, Fig.~\ref{fig4}(a). A non-conservative positional force $\beta r$ acts on the mass. With $u$ and $v$ representing the displacements in the direction of the two rotating coordinate axes, respectively, the system is governed by the equations \cite{SM68}
\ba{e1}
m\ddot{u}+\mu_1\dot{u}-2m\Omega\dot v+(k_1-m\Omega^2)u+\beta v&=&0,\nn\\
m\ddot{v}+\mu_2\dot{v}+2m\Omega\dot u+(k_2-m\Omega^2)v-\beta u&=&0.
\ea

In Fig.~\ref{fig4}(b) we show a numerically found surface of frequencies for the shaft with $m=1$ and $k_1=4$ in the absence of damping and non-conservative forces. The surface has four conical singularities corresponding to the doublets $\pm 2i$ at $\Omega=0$ and to the double zero eigenvalues at the critical speeds $\Omega=\pm2$. The cones in the subcritical speed range are near-vertically oriented while those at the critical speeds are near-horizontal. Consequently, for small stiffness detuning $\kappa$ the system is stable in the subcritical speed range and unstable by divergence in the vicinity of the critical speeds, where the bubbles of instability in the decay rate plots originate.

Addition of the non-conservative forces with $\beta=0.2$ and damping with $\mu_1=0.1$ and $\mu_2=0.2$ yields deformation of the conical surfaces with the apexes at $\Omega=0$ into the double coffee-filters. The real parts form the viaduct singular surfaces shown in Fig.~\ref{fig4}(c) and (d).
In the absence of damping ($\mu_{1,2}=0$) the gyroscopic system with the potential and non-conservative positional forces
cannot be asymptotically stable in accordance with the theorem of \citeasnoun{L75}. It is unstable almost everywhere in the space of parameters and can be only marginally stable on the set of measure zero in it. This is seen in Fig.~\ref{fig4}(c), which shows that the shaft is marginally stable at the points of the branch cuts, which form the set of measure zero, and unstable at all other points of the parameter plane.

\section{Example 2. A rotating circular string}

Consider a circular string of displacement
$W(\varphi, \tau)$, radius $r$, and mass per unit length $\rho$ that rotates with the speed $\gamma$ and
passes at $\varphi=0$ through a massless eyelet
generating a constant frictional follower force $F$ on the
string \cite{YH95}.
The circumferential tension $P$ in the string is constant;
the stiffness of the spring supporting the eyelet is $K$ and the damping coefficient of the viscous damper is $D$;
the velocity of the string in the $\varphi$ direction has constant value $\gamma r$.

With the non-dimensional variables and parameters
\be{s1}
t=\frac{\tau}{r}\sqrt{\frac{P}{\rho}},\quad w=\frac{W}{r},\quad \Omega=\gamma r\sqrt{\frac{\rho}{P}},
\quad k=\frac{Kr}{P},\quad \mu=\frac{F}{P}, \quad d=\frac{D}{\sqrt{\rho P}}
\ee
the substitution of $w(\varphi, t)=u(\varphi)\exp(\lambda t)$ into the governing equation and boundary conditions
yields the boundary eigenvalue problem \cite{YH95}
\be{s4}
Lu=\lambda^2 u+ 2\Omega\lambda u'-(1-\Omega^2)u''=0,
\ee
\be{s5}
u(0)-u(2\pi)=0,\quad u'(0)-u'(2\pi)=\frac{\lambda d+k}{1-\Omega^2}u(0)+\frac{\mu}{1-\Omega^2}u'(0),
\ee
where $'=\partial_{\varphi}$.
The boundary eigenvalue problem \rf{s4} and \rf{s5} depends on
the speed of rotation $(\Omega)$, and damping $(d)$, stiffness $(k)$, and
friction $(\mu)$ coefficients.

We note that the artificialness of the term, corresponding to the non-conservative positional forces, in the second of the boundary conditions \rf{s5},
was discussed in the literature, see, e.g. \cite{YH95,Ki08}. We keep it, however, to show how the degeneracy of this operator is seen in the eigenvalue surfaces.

\begin{figure}
\includegraphics[width=0.9\textwidth]{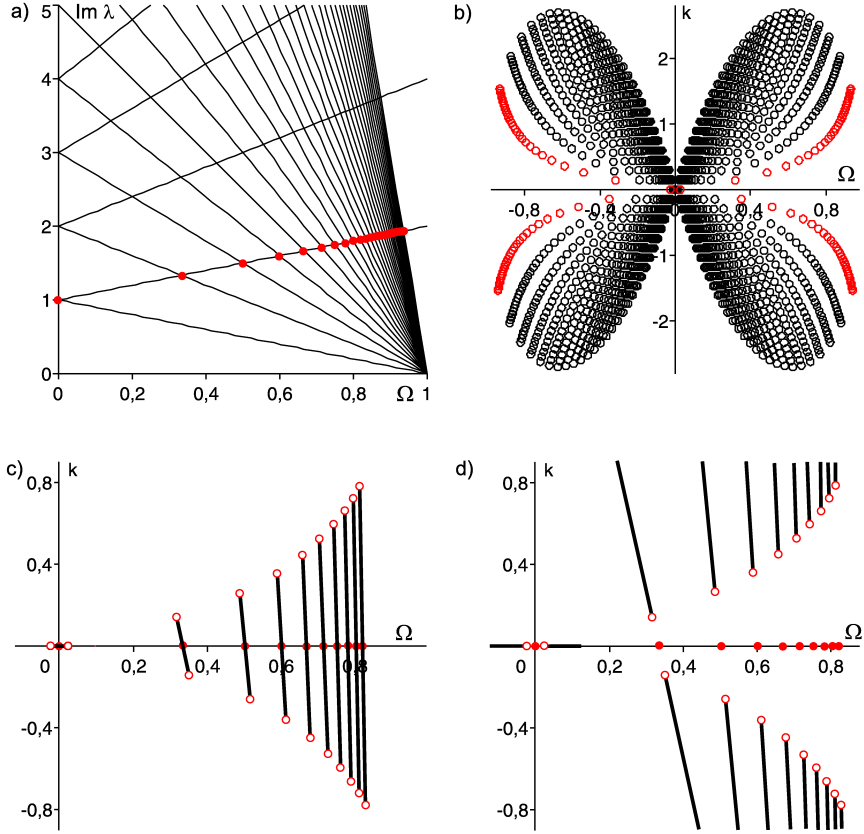}
\caption{\label{fig5} (a) The Campbell diagram of the unperturbed rotating string with red dots marking the nodes with $n=1$; (b) a butterfly distribution \rf{s16} of the exceptional points (open circles) in the subcritical speed range in the $(\Omega,k)$-plane when $\mu=0$ and $d=0.3$ (red open circles correspond to $n=1$); (c) projections of the branch cuts \rf{s17} of the coffee filters ${\rm Im}\lambda(\Omega,k)$ and the exceptional points for $n=1$; (d) projections of the branch cuts \rf{s17} of the viaducts ${\rm Re}\lambda(\Omega,k)$ and the exceptional points for $n=1$.  }
\end{figure}

For $d=0$, $k=0$, and $\mu=0$ the eigenvalue problem \rf{s4}, \rf{s5} has the eigenvalues $\lambda_n^{\varepsilon}=in(1+\varepsilon\Omega)$, $\lambda_m^{\delta}=im(1+\delta\Omega)$,
where $\varepsilon,\delta=\pm1$ and $n,m\in \mathbb{Z}-\{0\}$. In the $(\Omega,{\rm Im}\lambda)$-plane the branches intersect each other at the node
$(\Omega_0,\omega_0)$ with
\be{s11}
\Omega_0=\frac{n-m}{m\delta-n\varepsilon},\quad
\omega_0=\frac{nm(\delta-\varepsilon)}{m\delta-n\varepsilon},
\ee
where the double eigenvalue $\lambda_0=i\omega_0$ has two linearly independent eigenfunctions
\be{s13}
u_n^{\varepsilon}=\cos(n \varphi)-{\varepsilon}i\sin(n \varphi),\quad u_m^{\delta}=\cos(m \varphi)-\delta i\sin(m\varphi).
\ee
Intersections of the branch with $n=1$ and $\varepsilon=1$ and the branches with $m>0$ and $\delta<0$ in the subcritical range  $(|\Omega|<1)$ are marked in Fig.~\ref{fig5}(a) by red dots.

Taking into account that $\delta=-\varepsilon$ at all the crossings, excluding $(\Omega_0=\pm 1, \omega_0=0)$ where $\delta=\varepsilon$, we find approximation to the real and imaginary parts of the perturbed non-zero double eigenvalues \cite{Ki08}
\ba{s14}
{\rm Re}\lambda&=&-d\frac{n+m}{8\pi nm}\omega_0\pm\sqrt{\frac{|c|-{\rm Re}c}{2}},\nn \\ {\rm Im}\lambda&=&\omega_0+\varepsilon\frac{ n - m}{2}\Delta\Omega+\frac{n+m}{8\pi nm}k\pm\sqrt{\frac{|c|+{\rm Re}c}{2}},
\ea
where $\Delta\Omega=\Omega-\Omega_0$, and for the complex coefficient $c$ we have
\ba{s15}
{\rm Im}c&=&k\frac{2d\omega_0-\varepsilon\mu( n- m)}{16\pi^2nm}-2\left(\varepsilon\frac{ n + m}{2}\Delta\Omega+\frac{m-n}{8\pi nm}k\right)\left(\frac{\varepsilon}{4\pi}\mu-d\frac{m-n}{8\pi nm}\omega_0\right),\nn\\
{\rm Re}c&=&\left(\frac{\varepsilon n -\delta m}{2}\Delta\Omega+\frac{m-n}{8\pi nm}k\right)^2+\frac{k^2}{16\pi^2nm}
-\frac{\left[d(m+n)\omega_0\right]^2}{64\pi^2n^2m^2}.
\ea
Setting ${\rm Re}c=0$ and ${\rm Im}c=0$ we find the coordinates of the projections of the exceptional points of the surfaces ${\rm Re}\lambda(\Omega,k)$ and
${\rm Im}\lambda(\Omega,k)$ onto the $(\Omega,k)$-plane
\be{s16}
\Omega_{EP}=\Omega_0\pm\frac{\varepsilon}{8\pi nm}\frac{(m+n)d^2\omega_0^2}{\sqrt{nm(\mu^2nm+d^2\omega_0^2)}},~~
\kappa_{EP}=\pm \frac{d\omega_0(2\varepsilon \mu nm-d(m-n)\omega_0)}{2\sqrt{nm(\mu^2nm+d^2\omega_0^2)}}.
\ee
As in formulas \rf{cf1} the existence of the exceptional points \rf{s16} depends on the symplectic (Krein) signature of the intersecting branches, i. e. on the sign of $nm$, where $n,m \in \mathbb{Z}-\{0\}$. In the case of the rotating string all the crossings in the subcritical speed range $(|\Omega|<1)$ have definite Krein signature $(nm>0)$.
For those in the supercritical speed range $(|\Omega|>1)$ it is mixed with $nm<0$.
In the $(\Omega,\kappa)$-plane the exceptional points are situated on the line ${\rm Im}c=0$
\be{s17}
k=2\pi\varepsilon( n + m)\frac{2\varepsilon nm\mu-d\omega_0(m-n)}{d\omega_0(m^2+n^2)}\Delta\Omega.
\ee

\begin{figure}
\includegraphics[width=0.99\textwidth]{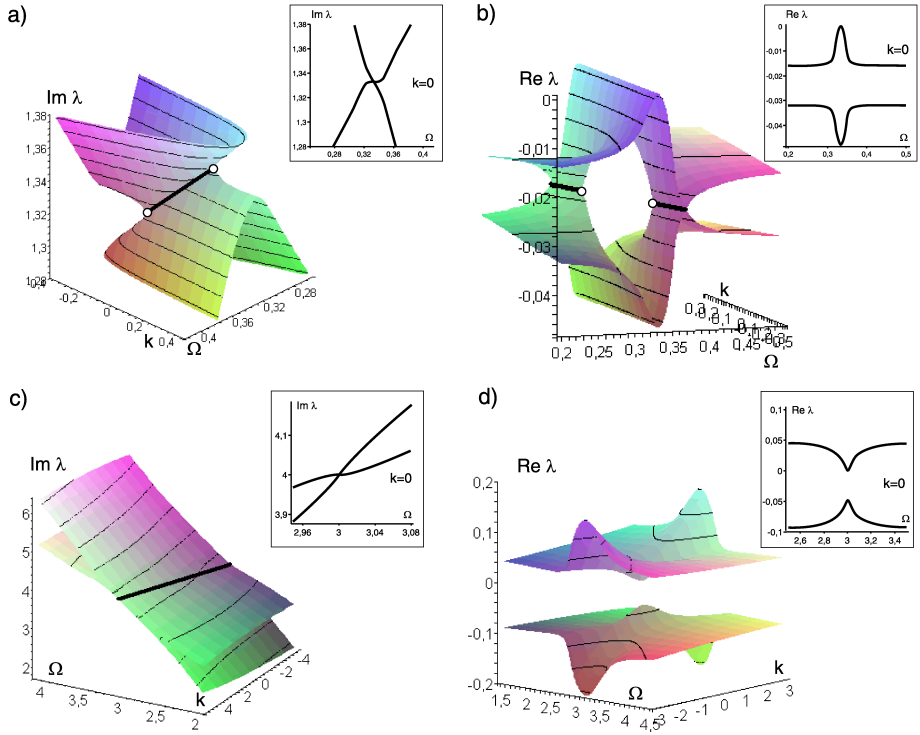}
\caption{\label{fig6} For $d=0.3$ and $\mu=0$: (a) the double coffee filter singular surface ${\rm Im}\lambda(\Omega,k)$ in the vicinity of the crossing $(n=1, m=2)$; (b) the viaduct surface ${\rm Re}\lambda(\Omega,k)$ corresponding  to the crossing $(n=1, m=2)$; (c) intersecting surfaces ${\rm Im}\lambda(\Omega,k)$ in the vicinity of the crossing $(n=1, m=-2)$  and (d) the corresponding non-intersecting surfaces ${\rm Re}\lambda(\Omega,k)$.}
\end{figure}

In Fig.~\ref{fig5}(b) we show the exceptional points \rf{s16} of the string passing through the eyelet with the damping coefficient $d=0.3$. The red open circles
correspond to the exceptional points born after the splitting of the diabolical crossings with $n=1$ and $\varepsilon=1$, which are shown in Fig.~\ref{fig5}(a) by the red dots. The exceptional points in the $(\Omega,\kappa)$-plane are distributed over a butterfly-shaped area, which preserves its form independently on the number of points involved. In comparison with the
numerical methods of \cite{J88,S07} our perturbation approach gives efficient explicit and interpretable expressions for the distribution of the exceptional points, for the branch cuts, and for the very eigenvalue surfaces.

In Fig.~\ref{fig5}(c) we plot the exceptional points originated after the splitting of the diabolical points with $n=1$ and $\varepsilon=1$ together with the projections of the branch cuts \rf{s17} of the double coffee filters ${\rm Im}\lambda(\Omega,k)$, which are shown by the bold lines. The corresponding projections of the branch cuts \rf{s17} of the viaducts  ${\rm Re}\lambda(\Omega,k)$ are presented in Fig.~\ref{fig5}(d). Only exceptional points originated after the perturbation of the doublets with $\Omega_0=0$ are situated on the $\Omega$-axis. This explains why damping creates a perfect bubble of instability for the doublets with $m=n$ and imperfect ones for the diabolical points with $m\ne n$ \cite{YH95,Ki08}.
Approximations \rf{s14} to the eigenvalue surfaces of a string with $\mu=0$ and $d=0.3$ are presented in Fig.~\ref{fig6} for different values of $n$, $m$, $\varepsilon$, and $\delta$. The smaller inclusions in Fig.~\ref{fig6} show the cross-sections of the surfaces by the plane $k=0$ for the convenience of comparing with the numerical data of \citeasnoun{YH95}. The results shown in Fig.~\ref{fig6} are in qualitative agreement with the developed theory for the equations \rf{i1} and \rf{i2} and perfectly agree with the numerical modeling.

\begin{figure}
\includegraphics[width=0.5\textwidth]{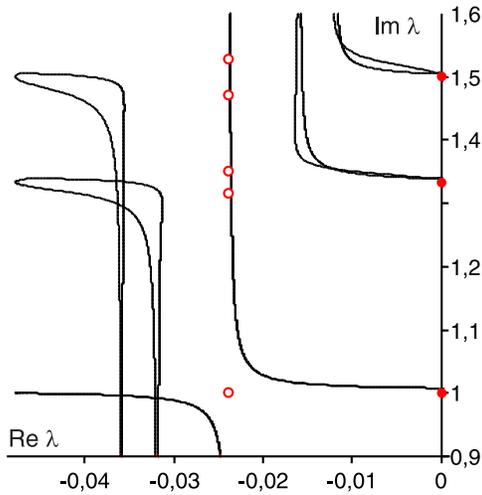}
\caption{\label{fig7} Exceptional points (open circles) with the parent diabolical points with $n=1$, $\varepsilon=1$ in the complex plane when $d=0.3$ and the trajectories $\lambda(\Omega)$ for $k=0.05$.}
\end{figure}

In Fig.~\ref{fig7} we show in the complex plane the parent diabolical points (red dots) and the corresponding exceptional points (open circles) whose locations are
\be{s18}
{\rm Re}\lambda_{EP}=- \frac{d}{4\pi},\quad {\rm Im}\lambda_{EP}=\frac{2nm}{n+m}\pm\frac{d}{4\pi}\frac{n-m}{\sqrt{nm}}.
\ee
In the engineering literature it was observed that the exceptional points ({\it strong modal resonances} \cite{D01}) are precursors to flutter instability because of their strong influence on the movement of eigenvalues in the complex plane.
Fig.~\ref{fig7} demonstrates the approximation of the `dynamics' of eigenvalues in the vicinity of the exceptional points, calculated by the formulas \rf{s14},
which is in a good qualitative agreement with the known numerical results \cite{STJ06}.

Finally, we notice that pure non-conservative positional perturbation $(d=0)$ causes degeneration of the eigenvalue surfaces.
Indeed, the line \rf{s17} reduces to $\Omega=\Omega_0$ and the two exceptional points merge into one at $\Omega_{EP}=\Omega_0$ and $\kappa_{EP}=0$. As a consequence, the central arch of the viaduct and the branch cut of the double coffee filter shrink to a single point. At this exceptional point the angle of crossing of the surfaces is zero in agreement with \cite{YH95,Ki08}. This degeneration visualizes the artificialness of the term related to the friction force in \rf{s15} that was already pointed out in the literature by physical arguments.

\section{Conclusion}

We found that in a weakly anisotropic rotor system \rf{i1} the branches of the Campbell diagram and the decay rate plots in the subcritical speed range are the cross-sections of the two companion singular eigenvalue surfaces. The double coffee filter and the viaduct are the imaginary and the real part of the unfolding of any double pure imaginary semi-simple eigenvalue at the crossing of the Campbell diagram with the definite symplectic (Krein) signature.
Generically, the structure of the perturbing matrices determines only the details of the geometry of the surfaces, such as the coordinates of the exceptional points and the spacial orientation of the branch cuts. It does not yield the qualitative changes irrespective of whether dissipative and circulatory perturbations are applied separately or in a mixture.
The two eigenvalue surfaces found unite seeming different problems on friction-induced instabilities in rotating elastic continua, because their existence does not depend on the specific model of the rotor-stator interaction and is dictated by the symplectic signature of the eigenvalues of the isotropic rotor and by the non-conservative nature of the forces originated at the frictional contact.
The double coffee filter singularity and its viaduct companion are true symbols of instabilities causing the wine glass to sing and the brake to squeal that connect these phenomena of the wave propagation in rotating continua with the physics of non-Hermitian singularities associated with the wave propagation in stationary anisotropic chiral media \cite{B04}.

\section*{Acknowledgements}
The work has been supported by the research grant DFG HA 1060/43-1.

\end{document}